# Effect of Misfit Strain in (Ga,Mn)(Bi,As) Epitaxial Layers on their Magnetic and Magneto-Transport Properties

K. Levchenko [1], T. Andrearczyk [1], J. Z. Domagała [1], J. Sadowski [1,2], L. Kowalczyk [1], M. Szot [1], T. Figielski [1] and T. Wosiński [1]

[1] *Institute of Physics, Polish Academy of Sciences, 02-668 Warszawa, Poland*
[2] *MAX-IV Laboratory, Lund University, P.O. Box 118, SE-221 00 Lund, Sweden*

Effect of misfit strain in the layers of (Ga,Mn)(Bi,As) quaternary diluted magnetic semiconductor, epitaxially grown on either GaAs substrate or (In,Ga)As buffer, on their magnetic and magneto-transport properties has been investigated. High-resolution X-ray diffraction, applied to characterize the structural quality and misfit strain in the layers, proved that the layers were fully strained to the GaAs substrate or (In,Ga)As buffer under compressive or tensile strain, respectively. Ferromagnetic Curie temperature and magneto-crystalline anisotropy of the layers have been examined by using magneto-optical Kerr effect magnetometry and low-temperature magneto-transport measurements. Post-growth annealing treatment of the layers has been shown to enhance the hole concentration and Curie temperature in the layers.

*levchenko@ifpan.edu.pl

## 1. Introduction

Highly mismatched ternary semiconductor compound Ga(Bi,As) has recently emerged as promising material for possible applications in a new class of photonic and spintronic devices. The replacement of a small fraction of As atoms by much larger Bi atoms in GaAs requires highly non-equilibrium growth conditions, such as low-temperature molecular-beam epitaxy (LT-MBE) at the substrate temperatures of 200–400°C [1, 2], i.e. far below 580°C, which is optimal for the MBE growth of GaAs. The band-gap energy of Ga(Bi,As) alloys decreases rapidly with increasing Bi content due to an interaction of Bi 6p bonding orbitals with the GaAs valence band maximum [3, 4]. Moreover, the replacement of As atoms by much heavier Bi atoms results, owing to a large relativistic correction to the GaAs band structure, in a strong enhancement of spin-orbit coupling, accompanied by a giant separation of the spin-split-off hole band [5, 6]. The increased spin-orbit coupling is especially favourable for spintronic materials where spin precession can be electrically tuned via the Rashba effect [7].

On the other hand, another GaAs-based ternary compound (Ga,Mn)As, in which a few percent of Ga lattice atoms have been substituted by Mn impurities, has become a prototype diluted ferromagnetic semiconductor, which exhibits spintronic functionalities associated with collective ferromagnetic spin ordering. Substitutional Mn ions in (Ga,Mn)As become ferromagnetically ordered below the Curie temperature owing to interaction with spin-polarized holes. The sensitivity of the magnetic properties, such as the Curie temperature and magnetic anisotropy, to the hole concentration allows for tuning those properties by post-growth annealing, photo-excitation or electrostatic gating of the (Ga,Mn)As layers [8, 9]. Moreover, appropriate nanostructurization of thin (Ga,Mn)As layers offers the prospect of taking advantage of magnetic domain walls in novel spintronic devices [10, 11].

In the present paper we report on an effect of misfit strain in thin epitaxial layers of (Ga,Mn)(Bi,As) quaternary compound, grown under either compressive or tensile strain, on their magnetic and magneto-transport properties.

## 2. Experimental details

We have investigated (Ga,Mn)(Bi,As) thin layers of 15 and 50 nm thicknesses, with 6% Mn and 1% Bi contents, grown by the low-temperature MBE technique at a temperature of 230°C on either semi-insulating (001)-oriented GaAs substrate or the same substrate covered with a 0.63-μm thick $In_{0.2}Ga_{0.8}As$ buffer layer. After the growth the samples were cleaved into two parts. One part of each sample was subjected to the low-temperature annealing treatment performed in air at the temperature of 180°C during 50 h. Annealing at temperatures below the growth temperature can substantially improve magnetic and transport properties of thin (Ga,Mn)As layers due to outdiffusion of charge- and moment-compensating Mn interstitials from the layers [12].

Both the as-grown and the annealed samples, containing the (Ga,Mn)(Bi,As) layers of 50 nm thicknesses, have been subjected to high-resolution X-ray diffraction (XRD) characterization at room temperature. Lattice parameters and misfit strain in the layers were investigated using reciprocal lattice mapping and rocking curve techniques for both the symmetric 004 and asymmetric $\bar{2}\bar{2}4$ Bragg reflections of Cu K$\alpha_1$ radiation. The thinner (Ga,Mn)(Bi,As) layers of 15 nm thicknesses, grown under the same conditions as the thicker ones, have been subjected to investigations of their magnetic and magneto-transport properties. Magnetic properties of the (Ga,Mn)(Bi,As) layers were examined using magneto-optical Kerr effect (MOKE) magnetometry. The MOKE experiments were performed both in longitudinal and polar geometries using He-Ne laser as a source of linearly polarized light with the laser spot of about 0.5 mm in diameter. The angle of incidence of light on the sample was about 30° for the longitudinal and 90° for the polar geometry. The standard lock-in technique with photo-elastic modulator operating at 50 kHz and a Si diode detector was used. Measurements of magnetic hysteresis loops were performed in the temperature range $T = 6-150$ K and in external magnetic fields up to 2 kOe applied in the plane of the layer and perpendicular to the layer for the longitudinal and polar geometry, respectively. Magneto-transport properties of the layers have been measured at liquid helium temperatures in samples of Hall-bar shape supplied with Ohmic contacts to the (Ga,Mn)(Bi,As) layers using a low-frequency lock-in technique, as described in our earlier paper [13].

### 3. Experimental results and discussion

High-resolution X-ray diffraction characterization of the investigated layers confirmed their high structural perfection and showed that the layers grown on GaAs substrate were pseudomorphically strained to the substrate under compressive misfit strain. The 004 diffraction patterns recorded from the layers displayed symmetric (Ga,Mn)(Bi,As) peaks with strong interference fringes, indicating homogeneous layer compositions and good interface quality. An addition of a small amount of Bi to the (Ga,Mn)As layers resulted in a distinct increase in their lattice parameter perpendicular to the layer plane and an increase in the in-plane compressive strain [14, 15]. On the other hand, the (Ga,Mn)(Bi,As) layers grown on the (In,Ga)As buffer layer, with distinctly larger lattice parameter, were pseudomorphically strained to the buffer under tensile misfit strain, as obtained from the $\bar{2}\bar{2}4$ reciprocal lattice maps, cf. Fig. 1. The annealing treatment applied to the (Ga,Mn)(Bi,As) layers resulted in a decrease in their lattice parameters and the strain. Taking into account the recent Rutherford backscattering spectrometry results by Puustinen et al. [16], which gave no evidence of Bi diffusing out of Ga(Bi,As) layers during annealing at temperatures of up to 600°C, we assume that the observed decrease in lattice parameters of the (Ga,Mn)(Bi,As) layers under the annealing treatment resulted mainly from outdiffusion of Mn interstitials from the layers. The in-plane misfit strain (lattice mismatch) values in the annealed (Ga,Mn)(Bi,As) layers, calculated from the XRD results, were $4.6 \times 10^{-3}$ (compressive strain) and $-9.7 \times 10^{-3}$ (tensile strain) for the layers grown on the GaAs substrate and on the (In,Ga)As buffer, respectively.

The longitudinal MOKE measurements were performed as a function of magnetic field applied along the main in-plane crystallographic directions: [100], [110] and $[\bar{1}10]$. The representative MOKE magnetization hysteresis loops of the (Ga,Mn)(Bi,As) layer grown on GaAs are presented in Fig. 2. The layers grown on GaAs displayed nearly rectangular hysteresis loops in the magnetic reversal measured with longitudinal MOKE and no magnetic reversal measured with polar MOKE, evidencing for the in-plane magnetization. Closer inspection of the MOKE hysteresis loops obtained under a magnetic field along the main in-plane crystallographic directions indicates easy magnetization axes along the in-plane ⟨100⟩ cubic directions and hard axes along two magnetically nonequivalent in-plane ⟨110⟩ directions, with the $[\bar{1}10]$ direction being magnetically easier than the perpendicular [110] one. Such a rather complicated magneto-crystalline anisotropy is characteristic of (Ga,Mn)As layers grown under compressive misfit strain [11, 17].

On the other hand, the (Ga,Mn)(Bi,As) layers grown on the (In,Ga)As buffer displayed clear hysteresis loops while measured in polar MOKE geometry, as shown in Fig. 3, and no magnetic reversal measured in longitudinal MOKE geometry. These results evidence for the easy magnetization axis along the [001] growth direction, characteristic of (Ga,Mn)As layers grown under tensile misfit strain [17]. The coercive field for the annealed (Ga,Mn)(Bi,As) layer grown on the (In,Ga)As buffer was 670 Oe at $T = 6$ K (cf. Fig. 3) and it was much larger than the coercive field of 75 Oe at the same temperature for the annealed (Ga,Mn)(Bi,As) layer grown on GaAs (cf. Fig. 2). From the analysis of temperature dependences of MOKE magnetization hysteresis loops we have determined the ferromagnetic Curie temperatures, $T_C$, in the investigated layers. The as-grown layers displayed the $T_C$ values of about 60 K, which have been enhanced as a result of the annealing treatment to the values of 100 K and 125 K for the layers grown under compressive and tensile misfit strain, respectively.

The electrical-transport properties of ferromagnetic materials are strongly affected by their magnetic properties resulting in an anisotropic magneto-resistance, which reflects the materials' magneto-crystalline anisotropy [18]. Magneto-transport properties of the (Ga,Mn)(Bi,As) layers have been measured as a function of magnetic field applied both along the main in-plane crystallographic directions and perpendicular to the layer. While sweeping the magnetic field up and down in the range of ±1 kOe, the layer resistance varied non-monotonously displaying double hysteresis loops caused by a rotation of the magnetization vector in the layers between equivalent easy axes of magnetization [13]. The magneto-transport results fully confirmed the in-plane and out-of-plane easy axes of magnetization in the layers grown under compressive and tensile strain, respectively. Moreover, those results clearly evidenced a significant increase in the hole concentration as a result of the annealing treatment, which was the main reason of the observed enhancement of the layer Curie temperature.

## 4. Summary and conclusions

Homogeneous layers of the (Ga,Mn)(Bi,As) quaternary diluted magnetic semiconductor have been grown by the low-temperature MBE technique on either GaAs substrate or the same substrate covered with a thick (In,Ga)As buffer layer. High-resolution X-ray diffraction characterization of the layers showed that in both cases they were grown pseudomorphically, under compressive or tensile misfit strain, respectively. Magnetic properties of the layers were examined with the MOKE magnetometry and low-temperature magneto-transport measurements. The obtained results revealed the in-plane and out-of-plane easy axis of magnetization in the layers grown under compressive and tensile misfit strain, respectively. Post-growth annealing of the layers, causing outdiffusion of self-compensating Mn interstitials, results in significant increase in the layer Curie temperature and the hole concentration. Incorporation of a small amount of Bi into the (Ga,Mn)As layers, which results in distinctly stronger spin-orbit coupling in the layers, does not markedly change their magnetic properties.

## Acknowledgment

This work was supported by the Polish National Science Centre under grant No. 2011/03/B/ST3/02457.

**Figure captions**

Fig. 1. Reciprocal lattice map of the annealed (Ga,Mn)(Bi,As)/(In,Ga)As/GaAs heterostructure for the $\bar{2}\bar{2}4$ XRD reflection where the vertical and horizontal axes are along the out-of-plane [001] and in-plane [110] crystallographic directions, respectively, in reciprocal lattice units. The solid and dashed lines denote the reciprocal lattice map peak positions calculated for pseudomorphic and fully relaxed layers, respectively.

Fig. 2. Magnetization hysteresis loops of the annealed (Ga,Mn)(Bi,As) layer grown on GaAs substrate recorded at various temperatures (written in the figure) with the longitudinal MOKE magnetometry under an in-plane magnetic field along the [100] crystallographic direction. The curves have been vertically offset for clarity.

Fig. 3. Magnetization hysteresis loops of the annealed (Ga,Mn)(Bi,As) layer grown on the (In,Ga)As buffer recorded at various temperatures (written in the figure) with the polar MOKE magnetometry under a magnetic field along the out-of-plane [001] crystallographic direction. The curves have been vertically offset for clarity.

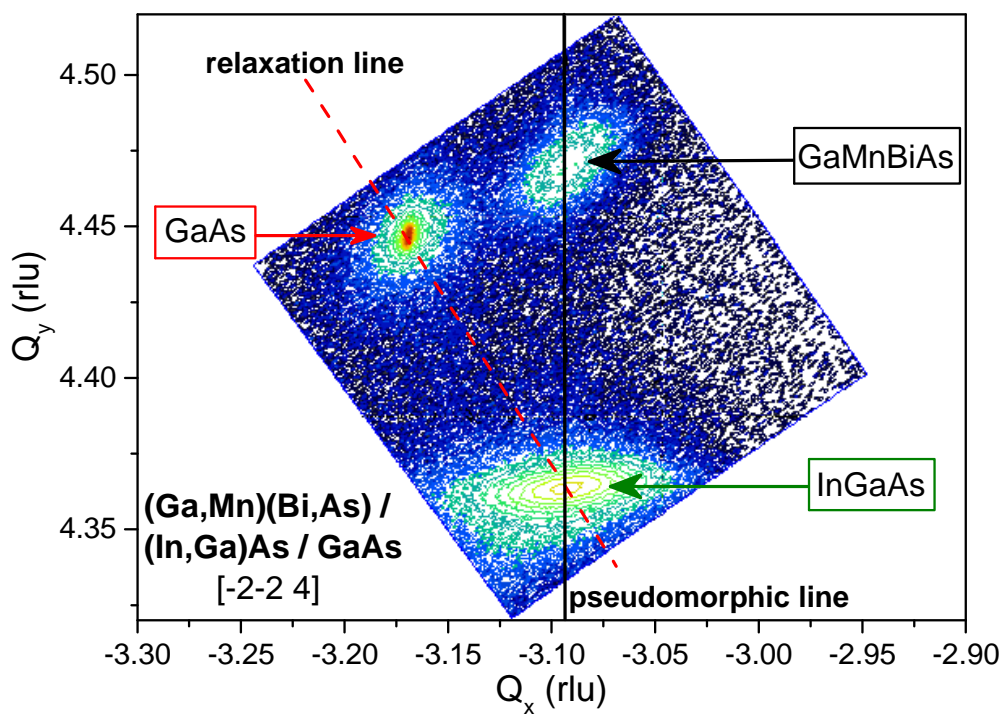

Fig. 1

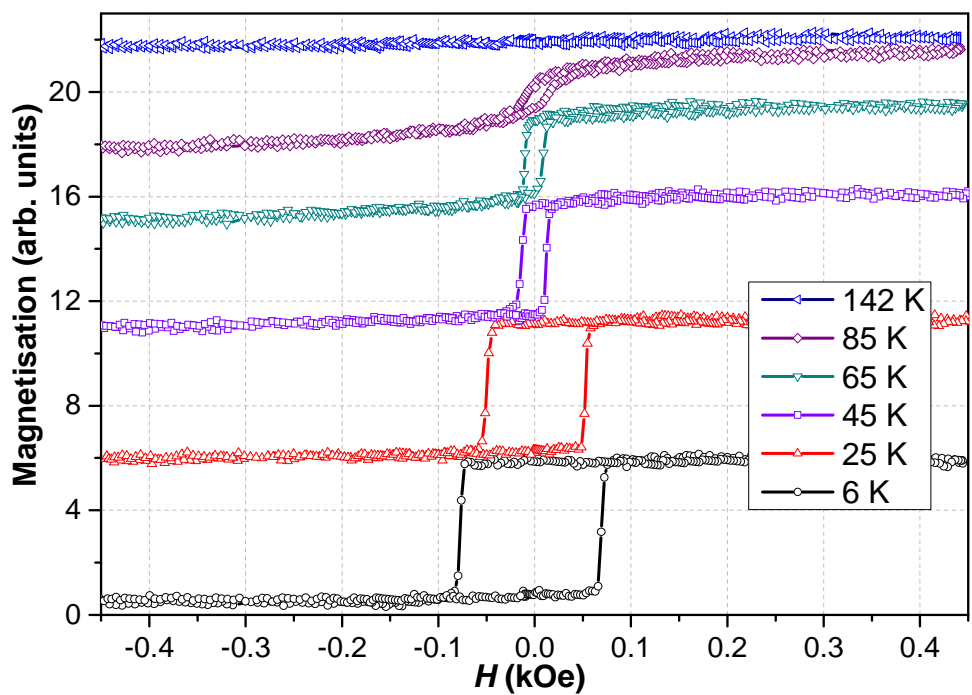

Fig. 2

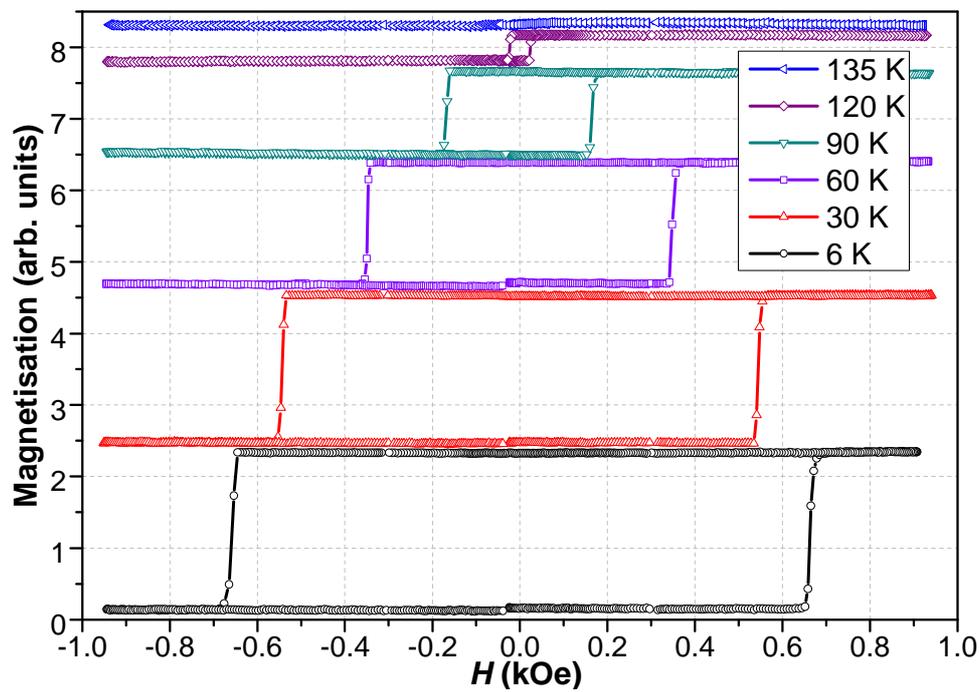

Fig. 3